\documentstyle[preprint,aps,epsfig]{revtex}
\begin{document}
\draft
\tighten
\preprint{JLAB-THY-01-4}
\title{Mesonic cloud contribution to the nucleon and $\Delta$ masses}
\author{D. J. Ernst}
\address{Department of Physics and Astronomy, Vanderbilt University,
Nashville, TN 37235\\
Jefferson Lab, 12000 Jefferson Avenue, Newport News, VA 23606}
\date{\today}
\maketitle
\begin{abstract}
Pion-nucleon elastic scattering in the dominant $P_{33}$ channel is examined in the  
model in which the interaction is of the form $\pi + N\leftrightarrow N,\,\Delta(1232)$. 
New expressions are found for the 
elastic pion-nucleon scattering amplitude which differ from existing formula both in 
the kinematics and in the treatment of the renormalization of the nucleon mass and 
coupling constant. Fitting the model to the phase shifts in the $P_{33}$ channel 
does not uniquely fix the parameters of the model. The cutoff for 
the pion-nucleon form factor is found to lie in the range $\beta = 750\pm350$ MeV/c. 
The masses of the nucleon and the $\Delta$ 
which would arise if there were no coupling to mesons are found to be
$m_{_N}^{(0)}= 1200\pm 200$ MeV and $m_\Delta^{(0)} = 1500\pm 200$ MeV. The difference 
in these bare masses, a quantity which would be accounted for by a residual gluon 
interaction, is found to be $\delta m^{(0)}=350\pm 100$ MeV.
\end{abstract}
\section{Introduction}
\label{formal}

One of the most challenging problems \cite{NIbook} facing contemporary physics is the 
understanding of Quantum Chromodynamics (QCD) in the region of confined quarks. 
Lattice QCD has made great progress in its ability to calculate physical quantities 
but it remains far distant from being able to calculate something like the nucleon 
wave function. Models of the nucleon and the excited baryons are thus necessary. It is 
possible that the relationship between lattice gauge calculations and nature may initially 
proceed through phenomenological models, chiral expansions, and effective Lagrangians that 
produce parameters that are more 
amenable to lattice calculations than might be the measurable quantities themselves. 

Models of baryons based on confined quarks\cite{Isgur,LYG96,MIT,CBM} are 
capable of producing a number of the measured properties of the baryons. We are here 
interested in a specific question. How do you model the pion (and other meson) cloud 
contributions \cite{CBM,picloud} to the structure of the nucleon and the 
$\Delta(1232)$? We take the approach that the meson-nucleon interaction cannot be 
treated perturbatively. The results we find are consistent with this assumption. There 
exists very accurate data \cite {SAID} for pion-nucleon scattering. In understanding 
the pion-nucleon system, we believe that this data must be used as a constraint. The 
elastic scattering pion-nucleon amplitude contains the nucleon pole which occurs at 
a pion mass below elastic threshold. The residue of the pole is the 
square of the physical nucleon wave function. Thus the mesonic cloud contribution to 
the nucleon is intimately related to the scattering data, just as the scattering wave 
function from a potential is not independent of the bound state wave functions for 
that same potential. The question is how to use the pion-nucleon data to constrain 
models of the pionic cloud of the single nucleon and the $\Delta$?

\begin{figure}
\epsfxsize=6.5in  
\epsfbox{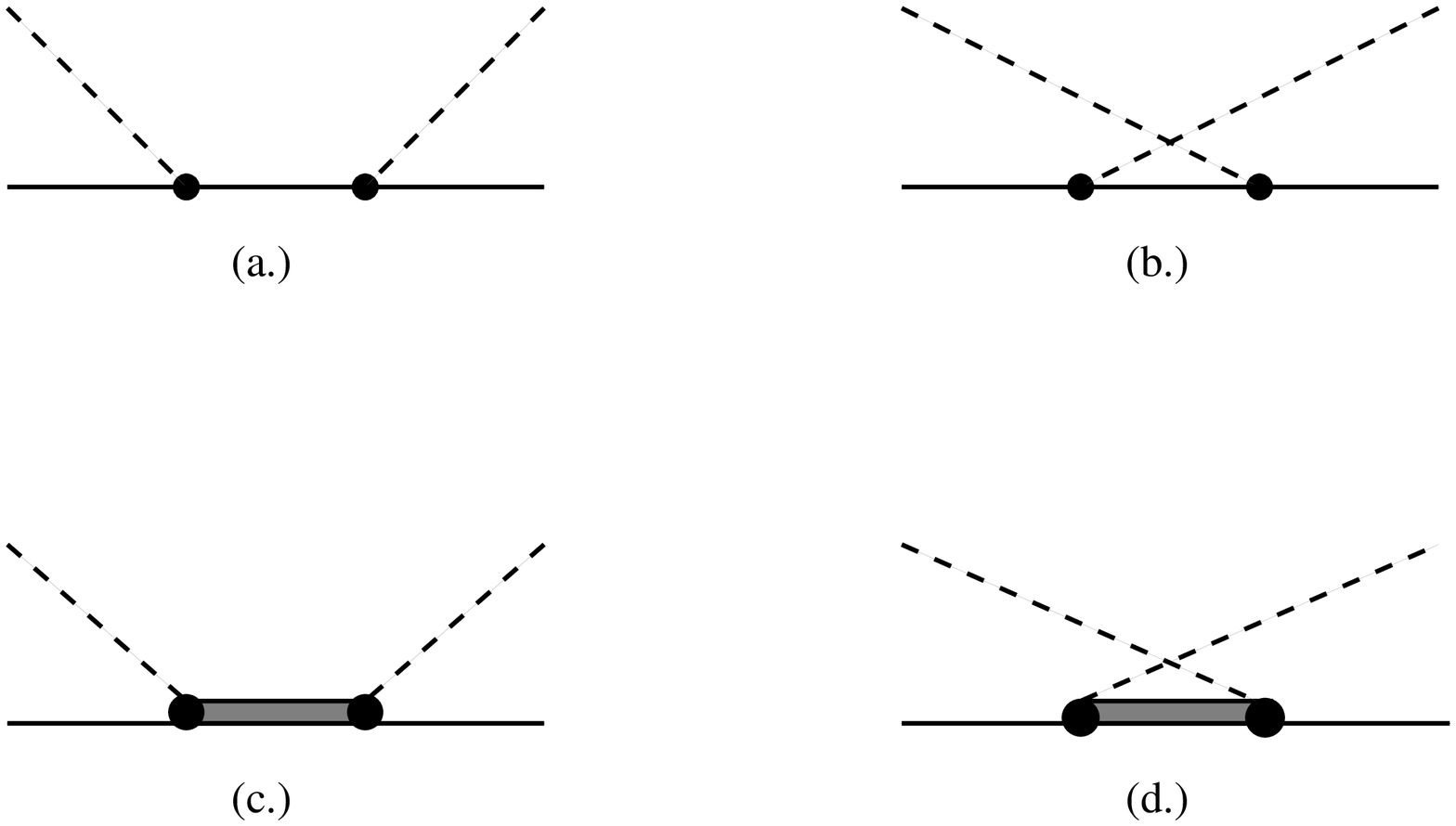} 
\vspace {25pt} 
\caption
{Driving terms for pion-nucleon scattering: a.) the direct nucleon term, b.) the 
crossed 
nucleon term, 
c.) the direct $\Delta$ term and d.) the crossed $\Delta$ 
term. The combined model for scattering 
in the $P_{33}$ channel developed here includes b.) and c.) as driving terms for the 
linear equations.}
\label{xdir}
\end{figure}

A first step in answering this question is presented here. We adopt a model in which 
the coupling is of the form $\pi+N\leftrightarrow N,\,\Delta$. We then investigate how 
to calculate pion-nucleon scattering given this model of the interaction.  There exists a 
large number \cite{CBM,pot,LM85,chlw,DJE78b,DJE90,RJM81,JBC73,tshl,MGF95,FG93,BCP91,CS94} of 
models of pion-nucleon scattering. We require a model which contains the pion-nucleon 
pole term, both because such a term has long been known to be physically present in the 
amplitude and because this is how we will be able to extract information on the nucleon 
itself from the model. We believe we should begin with the spin-isospin channel which is 
dominant at low energies, the $P_{33}$ channel. In this channel, we assume that the 
dominant physics arises from the crossed nucleon pole, Fig.~\ref{xdir}b, and the 
direct $\Delta$ production, Fig.~\ref{xdir}c, utilized as the lowest order driving 
terms of the theory. The model is then conceptually the same as the Cloudy Bag Model 
\cite{CBM}. Expressions for the scattering amplitude within this model have been 
derived in \cite{CBM} and \cite{RJM84}. We find here that a more complete treatment of 
the renormalization of the nucleon mass and the pion-nucleon coupling constant 
provides a new result which when fit to the data gives qualitatively different results
from these previous works. 

The model is formulated in such a way as to produce an interesting piece of  
information concerning the structure of the nucleon and the $\Delta$. The physical  
picture of the nucleon that underlies the model is that there is a core composed of  the 
valence quarks surrounded by a mesonic cloud. Within the model, one can calculate  the 
mass of a baryon in the absence of the coupling to the mesons.  This 
mass, here referred to alternately as the bare mass or unrenormalized mass, is a  property 
of the valence quarks only. Symmetry arguments should apply well to the  valence quarks, 
which we assume to have a reasonably simple structure, and not so well  to the physical
particles, given we find they have significant mesonic cloud contributions.  Thus the
process of modeling the mesonic cloud and removing its contribution to  baryonic
properties can provide insight into the simpler valence quark structure.

We here address the question, given the model interaction, how can one best solve for 
elastic pion-nucleon scattering. There is a second important question which we do not 
address. How does one generate the underlying model of the pion-nucleon 
interaction from meson-quark or quark-quark interactions. For example, in the Cloudy Bag 
Model \cite{CBM}, the pion-nucleon 
coupling is generated by coupling the pion to the valence quarks at the surface of an 
MIT bag \cite{MIT} in such a way as to preserve chiral symmetry. Such a model is of 
the type we envision underlying this work. The underlying model of the coupling 
produces the form factor for the $\pi + N\leftrightarrow N,\,\Delta$ interactions. Pion 
nucleon scattering does not seem to be sensitive to the exact function chosen for the 
form factor, so we defer discussion of the source of the pion-nucleon coupling as a 
separate problem, and treat the form factor as a phenomenological quantity whose range 
is to be determined from data.

The formalism developed here has a finite mass target, uses invariant phase space and 
normalizations, and works with 
the invariant amplitude that is free of kinematic singularities. In Sec.~\ref{sec2} we provide 
expressions for quantities needed to develop the model --- the model interaction, the 
approximate crossing relation used, and the pole terms of the pion-nucleon scattering 
amplitude. In Sec.~\ref{CLL} we 
review separately the Chew-Low model, where the coupling is $\pi+N\leftrightarrow N$, 
and the Lee model, where the coupling is $\pi + N\leftrightarrow \Delta$. A 
relationship between the models is found which leads us in Sec.~\ref{comb} to a new 
solution for the scattering amplitude when both interactions are present. In 
Sec.~\ref{results}, the parameters of the model are fit to the $P_{33}$ pion-nucleon 
phase shifts. In the Conclusions, the results of this work are summarized and thoughts 
on future work are presented.

\section{Model interaction and crossing relation}
\label{sec2}

We first need a model interaction, an approximate crossing relation, and expressions for the 
pole terms in the pion-nucleon scattering amplitude. For pion-nucleon 
scattering, we propose using an interaction composed of three 
point functions, $\pi+N\leftrightarrow N^*$. In the $P_{33}$ channel that we will examine 
here, the dominate physics arises from  $N^*=N$, $\Delta$. Anticipating future 
applications to the higher baryon resonances, we will for now treat $N^*$ as any 
baryon state. A three point coupling is given by
\begin{equation} 
H_I^i= \int
\frac{d^3p'}{2E_{p'}}\frac{d^3p}{2E_{p}}\frac{d^3k}{2\omega_k}\, \langle\,{\bf
p}'\,\vert\,H_I^i\,\vert\,{\bf p},{\bf k}\,\rangle\, b^{i\,\dagger}_{{\bf
p}'}\,b^{^N}{\bf p}\,a_{\bf k} + h.c. 
\label{ham} 
\end{equation} 
with $i$ labeling the the $N^*$ baryon, $b^{i\,\dagger}_{\bf p'}$ the creation operator for 
baryon $i$ with momentum ${\bf p'}$, $b^{^N}_{\bf p}$ the destruction operator for a nucleon 
with momentum $\bf p$, $a_{\bf k}$ the destruction operator for a pion of
momentum ${\bf k}$, and $h.c.$ the hermitian conjugate of the previous term. We
reintroduce spin and isospin labels, and use Lorenz covariance, rotational 
invariance,
and isospin invariance, to write the interaction as 
\begin{eqnarray} 
\langle\,{\bf p}',j,j_3,\tau,\tau_3\,\vert\,H_I^i\,\vert\,
{\bf p},s^N_3,\tau^N_3; {\bf k},\tau^\pi_3\,\rangle
&=&2E_{p'}\,\delta(  {\bf p}'-{\bf p}-{\bf k})\nonumber\\  &&
i f^i_{j\tau\ell}\,C(\frac{1}{2},\tau^N_3,1,\tau^\pi_3;\tau,\tau_3) {\cal  Y}_{
\ell}^{jj_3}(\theta_q,\phi_q)^*\,q^\ell v^{j\tau}_\ell (q^2)\,\,, 
\label{ham2}
\end{eqnarray}
with $j$ and $\tau$ the spin and isospin of the $N^*$ and 
where 
\begin{equation} {\cal  Y}_{
\ell}^{jj_3}(\theta_q,\phi_q) =
\sum_{m,s^N_3}C(\ell,m,\frac{1}{2},s^N_3;jj_3) Y_{\ell m}(\theta_q,\phi_q)\,\,.
\label{Ylm} 
\end{equation} 
Note the factor $2E_{p'}$ which accompanies the momentum conserving $\delta$-function 
to insure covariance, and the explicit constructions introduced to maintain rotational
invariance and isospin invariance. For our model we take two terms in the interaction, 
one that couples to a particle with $j=1/2$,
$\tau=1/2$ and one with $j=3/2$, $\tau=3/2$, providing a coupling to the nucleon and 
to the $\Delta$ respectively. The momentum $q$ is defined as the momentum of the pion in
the reference frame where the total momentum is zero, ${\bf p}+{\bf k}=0$. The factor
$q^\ell$ is incorporated to produce the correct threshold behavior. For the coupling 
$\pi+N\leftrightarrow N$, this may be written in a more familiar form by using
\begin{equation}
q\sum_{m,s_3^N}C(1,m,\frac{1}{2},s_3^N;\frac{1}{2},j_3)Y^*_{\ell m}(\theta_q,\phi_q)=
-\frac{1}{\sqrt{4\pi}}\langle\,\frac{1}{2},j_3\,\vert\,\vec \sigma\cdot{\bf
q}\,\vert\,s_3^n\,\rangle\,\,.
\label{sigdq}
\end{equation}
The construction given in Eq.~\ref{ham2} is general and can be used for any value of
the spin $j$ and isospin $\tau$ of the intermediate baryon. The construction of the
state $\vert\,{\bf p},s_3^N;{\bf k}\,\rangle$ and the definition of the state
$\vert\,{\bf q}, s^N_3\,\rangle$ including Wigner spin precession, which we do not include 
here, is described in detail 
in Ref.~\cite{Gieb} for a spin 1/2 particle and in \cite{DVA92} for particles of 
arbitrary spin.  

The pion-nucleon amplitude will contain the direct pion-nucleon pole, Fig.~1a, given by
\begin{equation}
\langle\,q'\,\vert\,t_\alpha^{pole}\,\vert\,q\,\rangle=\delta_{\alpha,1}\lambda_\alpha
\frac{v_{_N}(q')\,v_{_N}(q)}{W_q-m_{_N}}
\label{poled}
\end{equation}
The subscript $\alpha$ is an abbreviation for $j,\,\tau,\,\ell$. The residue of the nucleon 
pole 
term in the $P_{11}$ 
channel is related to the conventional definition of the pion-nucleon coupling constant 
$f^2_{_{\pi NN}}$ by
$\lambda_1=12\,(m_{_N}/m_\pi)\,f^2_{_{\pi NN}}$.

In addition to the the direct nucleon pole term, there will also be crossed nucleon pole 
terms, Fig.~1b. These crossed nucleon pole terms are $U$-channel singularities while the direct 
term is an $S$-channel singularity. In dynamic models, it is very difficult \cite{RJM81,SBJ}
to work with 
the crossed channels as $U$-channels. We will here approximate the $U$-channel crossed 
terms by an $S$-channel singularity. The simplest approximation is
\begin{equation}
\langle\,q\,\vert\,t_\alpha(-W+2m_{_N})\,\vert\,q'\,\rangle = \sum_\beta 
A_{\alpha\beta}\,\langle\,q'\,\vert\,t_\beta(W)\,\vert\,q\,\rangle\,\,,
\label{cross}
\end{equation}
with the crossing matrix given by
\begin{equation}
A_{\alpha\beta}=\frac{1}{9} 
       \,\pmatrix{
       1&-8&-8&16\cr
       -2&-1&8&4\cr
       -2&8&-1&4\cr
       4&4&4&1\cr}\,\,\,.
\label{cmat}
\end{equation}
If we apply relationship \ref{cross} to the direct nucleon pole term in Eq.~\ref{poled} to 
generate 
the crossed nucleon pole terms, we find for the total
\begin{eqnarray}
\langle\,q'\,\vert\,t_\alpha^{pole}\,\vert\,q\,\rangle=\delta_{\alpha 1}\lambda_\alpha\,
  \frac{q'q\,v_{_N}(q')\,v_{_N}(q)}{W_q-m_{_N}}&+&\lambda_\alpha^x\,
  \frac{q'q\,v_{_N}(q')\,v_{_N}(q)}{-W_q+m_{_N}}\,\,, 
\label{polex}
\end{eqnarray}
with $\lambda_\alpha^x$ given by $\{1/9, -2/9, -2/9, 4/9\}\,\lambda_1$, for 
$\alpha=1,4$ representing the $P_{11}$, $P_{13}$. $P_{31}$, and $P_{33}$ channels.

\section{Chew-Low and Lee models}
\label{CLL} 

Before investigating the model with both couplings, $\pi+N\leftrightarrow N$ and 
$\pi+N\leftrightarrow \Delta$, we examine models where only one coupling is present. By 
investigating these, particularly how each model handles the renormalization of the 
nucleon mass, we will learn how to solve the combined model. 
The Lee model \cite{Lee} consists of choosing an interaction of the form 
$\pi+N\leftrightarrow \Delta$. We will also need to consider the case where the coupling is 
$\pi+N\leftrightarrow N$ and thus use $N^*$ to represent either $N$ or $\Delta$. The second 
order diagram is of the form of an energy-dependent separable potential, 
\begin{equation}
\langle\,q'\,\vert\,{\cal V}^{\, eff}_\alpha\,\vert\,q\,\rangle=
\lambda^{(0)}_\alpha\,\frac{q'q\,v_{_{N^*}}(q')\, 
v_{_{N^*}}(q)}{W_q-m^{(0)}_{_{N^*}}},
\label{leedrive}
\end{equation}
and serves as a driving term for the linear Lippman-Schwinger equation.
We have attached superscript zeros to the coupling constant and the mass of the 
$N^*$ to remind us that these are not renormalized quantities. We also examine the 
case where the second order term is of the form of a crossed Lee type interaction. 
From Eq.~\ref{cross}, this would be
\begin{equation}
\langle\,q'\,\vert\,{\cal V}^{\, eff}_\alpha\,\vert\,q\,\rangle=
\lambda^{x(0)}_\alpha\,\frac{q'q\,v_{_{N^*}}(q')\, 
v_{_{N^*}}(q)}{-W_q+2m_{_N}-m^{(0)}_{_{N^*}}}
\,\,\,,
\label{crossdrive}
\end{equation}
with $\lambda_\alpha^{x(0)}$ calculated from $\lambda_\alpha^{(0)}$ using Eq.~\ref{cross}.
Since these are of the form of an energy dependent separable potential, the solution 
for the scattering matrix follows by inserting the effective potential into the 
Lippman-Schwinger equation.
\begin{equation}
\langle\,q'\,\vert\,t_\alpha(W_q)\,\vert\,q\rangle=
\langle\,q'\,\vert\,{\cal V}^{\,eff}_\alpha(W_q)\,\vert\,q\,\rangle+\int \frac{q'' 
dq''}{4E_{q''}\omega_{q''}}\frac{\langle\,q'\,\vert\,{\cal 
V}_\alpha^{\,eff}(W_q)\,\vert\,q''\rangle\,\langle\,q''\,\vert\,t_\alpha(W_q)
\,\vert\,q\,\rangle}{W_q-W_{q''}+i\eta}\,\,.
\label{LpSw}
\end{equation}
The phase space factor arises from the use of invariant normalizations and working with the 
invariant amplitude. This equation is also known \cite{VGK81} as the Kadyshevski equation.

Parameterize $\langle\,q'\,\vert\,t_\alpha (W_q)\,\vert\,q\,\rangle$ by
\begin{equation}
\langle\,q'\,\vert\,t_\alpha (W_q)\,\vert\,q\,\rangle=
 \lambda^{(0)}_\alpha \,q'q\, v_{_{N^*}}(q')\,v_{_{N^*}}(q)/{\cal D}^L_\alpha 
(W_q)\,\,\,,
\label{leenum}
\end{equation}
with $\lambda^{(0)}_\alpha$ replaced by $\lambda^{x(0)}_\alpha$ if the driving term is 
the crossed term, Eq.~\ref{crossdrive}.
The result for the denominator function ${\cal D}^L_\alpha (W_q)$ is, for the direct 
driving term of Eq.~\ref{leedrive},
\begin{equation}
{\cal D}^L_\alpha (W_q)=W_q-m^{(0)}_{_{N^*}}-\lambda^{(0)}_\alpha\,\int\,\frac{q''^2      
               \,dq''}{4E_{q''}\omega_{q''}}\frac{q''^2\, 
v^2_{_{N^*}}(q'')}{W_q-W_{q''}+i\eta}\,\,\,,
\label{leeden}
\end{equation}
or for the crossed driving term of Eq.~\ref{crossdrive}
\begin{equation}
{\cal D}^L_\alpha 
(W_q)=-W_q+2m_{_N}-m^{(0)}_{_{N^*}}-\lambda^{x(0)}_\alpha\,\int\,\frac{q''^2      
               \,dq''}{4E_{q''}\omega_{q''}}\frac{q''^2\, 
v^2_{_{N^*}}(q'')}{W_q-W_{q''}+i\eta}\,\,\,.
\label{crossden}
\end{equation}

The question we need to address is what happens if the intermediate state, the $N^*$, 
is actually the nucleon itself. For the remainder of this section, we set $N^*= N$. 
In 
this case we would rewrite the results in terms of the physical, i.e. renormalized, 
nucleon mass. The direct and crossed nucleon pole terms, Eq.~\ref{polex}, arise from 
a zero of ${\cal D}^L_\alpha (W_q)$ at 
$W_q=m_{_N}$, or ${\cal D}^L_\alpha (W_q=m_{_N})=0$. This gives
\begin{equation}
m_{_N}^{(0)}=m_{_N}-\lambda_\alpha^{(0)}\, \int\,\frac{q''^2      
               \,dq''}{4E_{q''}\omega_{q''}}\frac{q''^2\, 
v^2_{_N}(q'')}{m_{_N}-W_{q''}}\,\,\,,
\label{renormm}
\end{equation}
and the {\it same} result (with $\lambda^{(0)}_\alpha$ replaced by 
$\lambda^{x(0)}_\alpha$) for the crossed driving term. 
If we substitute Eq.~\ref{renormm} into Eqs.~\ref{leeden} and \ref{crossden} to 
eliminate the unrenormalized nucleon mass, we find
\begin{equation}
{\cal D}^L_\alpha 
(W_q)=W_q-m_{_N}-(W_q-m_{_N})\,\lambda^{(0)}_\alpha\,\int\,\frac{q''^2      
               \,dq''}{4E_{q''}\omega_{q''}}\frac{q''^2\, 
v^2_{_N}(q'')}{(W_{q''}-m_{_N})(W_q-W_{q''}+i\eta})\,\,\,,
\label{denmren}
\end{equation}
for the direct driving term, and for the crossed driving term find
\begin{equation}
{\cal D}^L_\alpha 
(W_q)=-W_q+m_{_N}-(W_q-m_{_N})\,\lambda^{(0)}_\alpha\,\int\,\frac{q''^2      
               \,dq''}{4E_{q''}\omega_{q''}}\frac{q''^2\, 
v^2_{_N}(q'')}{(W_{q''}-m_{_N})(W_q-W_{q''}+i\eta})\,\,\,.
\label{denmcross}
\end{equation}
If we now make a change in notation, and define a coupling constant, 
$\tilde\lambda^{(0)}_\alpha$, by
\begin{equation}
\tilde\lambda^{(0)}_\alpha=\left\{
\begin{array}{c}\lambda^{(0)}_\alpha\\
-\lambda^{x(0)}_\alpha\end{array}
\right. \,\,\,,
\label{newcoup}
\end{equation}
then both cases, the direct and crossed driving terms, can be accommodated by using 
Eq.~\ref{denmren} with $\tilde\lambda_\alpha^{(0)}$ as the 
coupling constant. The minus that arises from the crossed diagram propagator has been 
absorbed into the coupling constant for notational convenience.

Finally, we identify the residue of the nucleon pole as the renormalized coupling 
constant, $\tilde\lambda_\alpha$. This implies 
\begin{equation}
\frac{1}{\tilde\lambda_\alpha}-\frac{1}{\tilde\lambda^{(0)}_\alpha}=
    \int\,\frac{q''\,dq''}{4E_{q''}\omega_{q''}}
    \frac{q''^2\,v(q'')}{(W_{q''}-m_{_N})^2}\,\,\,.
\label{conr}
\end{equation}
Substituting this back into Eqs.~\ref{leenum} and \ref{denmren} gives
the scattering matrix,
\begin{equation}
\langle\,q'\,\vert\,t_\alpha\,\vert\,q\,\rangle=
 \tilde\lambda_\alpha \,q'q\, v_{_N}(q')\,v_{_N}(q)/{\cal D}^{CL}_\alpha (W_q)\,\,\,,
\label{CLnum}
\end{equation}
with ${\cal D}^{CL}_\alpha(W_q)$ given by
\begin{equation}
{\cal D}^{CL}_\alpha (W_q)=(W_q-m_{_N})\,\left( 1-\tilde\lambda_\alpha                 
    \,(W_q-m_{_N})\int\,
   \frac{q''^2\,dq''}{4E_{q''}\omega_{q''}}
   \frac{1}{(W_{q''}-m_{_N})^2}                                                           
   \frac{q''^2\,v^2_{_N}(q'')}{W_q-W_{q''}+i\eta}\right) \,\,\,.
\label{CLden}
\end{equation}
This is the result for the Chew-Low model \cite{CL} in the no crossing approximation 
generalized for a finite nucleon 
mass. What we have found is that the Lee model, Eq.~\ref{leedrive}, and its crossed 
generalization, Eq.~\ref{crossdrive}, are equivalent to the Chew-Low model if the 
intermediate state in the Lee model is taken to be the nucleon. The relation of the 
Lee model to the Chew-Low model with a direct driving term was first noticed in 
Ref.~\cite{DJE90}. 
The generalization here to the crossed driving term is important as it will be needed in 
the next section. The Lee models, direct and crossed,  are written naturally in terms 
of the unrenormalized mass and coupling constant while the Chew-Low result is the 
equivalent written in terms of renormalized quantities.

It is interesting to note that the two Lee models, direct and crossed, differ in form when 
written in terms of unrenormalized quantities, but produce the same algebraic results when 
written in 
terms of renormalized quantities. Even when written in terms of renormalized 
quantities, however, the direct and crossed models are not equivalent. For the crossed 
driving term, the coupling constant $\tilde\lambda_\alpha$ is negative; it has been 
redefined to absorb the minus sign from the crossed propagator for the purpose of 
giving an algebraic similarity of the two models.

The renormalization of the nucleon mass, however, is the same for the two models when 
written in terms of the coupling constants $\lambda_\alpha^{(0)}$ or 
$\lambda_\alpha^{x(0)}$. This is important as this relation maintains for both cases 
the physical requirement that $m_{_N}^{(0)}>m_{N}$, i.e. the addition of a degree of 
freedom, here the pion-nucleon channel, lowers the energy of a state.

\section{Combined Model}
\label{comb}

\begin{figure}
\epsfxsize=6.5in  
\epsfbox{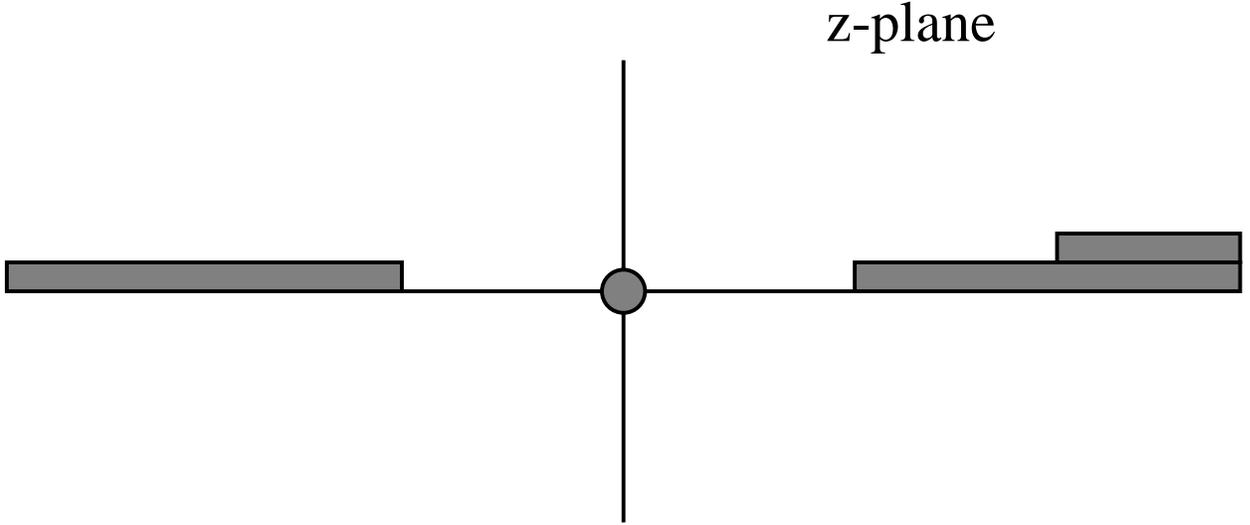}
\vspace {25pt} 
\caption[]
{Analytic structure of the pion-nucleon scattering 
amplitude 
in the complex $W_q\rightarrow z$ plane. The cut along the right-hand axis is composed of 
an 
elastic scattering contribution which starts at an energy of $m_{_N}+m_\pi$ together with 
an 
inelastic contribution starting at the inelastic threshold. There is a pole at $z=m_{_N}$ 
as given in Eq.~\protect\ref{polex}, and a 
cut 
along the left-hand axis, the crossed cut. In the model developed here, the left-hand cut 
is approximated by an increase in the residue of the pole term.}
\label{anal}
\end{figure}

We now return to the question of solving for the scattering amplitude for an 
interaction which contains both a $\pi + N \leftrightarrow N$ and a $\pi + N 
\leftrightarrow \Delta$. We limit the problem to the ``no crossing'' approximation. This 
approximation includes crossed terms through second order and their iterates. It is best 
understood in terms of the Low equation \cite{low} where crossing symmetry is manifest. We 
can understand why dropping the crossed term is a reasonable approximation, even 
though its contribution \cite{RJM82} to the scattering is not negligible. Examine the 
analytic structure of the pion-nucleon amplitude in the complex $W_q$ plane.  We 
picture this structure in Fig.~\ref{anal}, where we have employed the approximate crossing 
relation of Eq.~\ref{cross}. The no crossing approximation that we are using sets the left-hand 
cut to zero {\it and} compensates by increasing the residue of the nucleon pole. The 
physics we are examining is given by the scattering amplitude evaluated with the 
complex energy approaching the right-hand cut from above. In this region, the energy 
dependence of the actual nucleon pole term plus the crossing cut can be reasonably 
approximated by a pole with a modified residue. This approach does, however, preclude 
the use of the physical pion-nucleon coupling constant in the model.

We begin with a combination of Lee model driving terms, 
Eqs.~\ref{leedrive} and \ref{crossdrive}.  In the no crossing approximation, the 
scattering amplitude for a single interaction requires the solution of a linear 
equation. The solution for the scattering amplitude for an interaction which is the 
sum of the two terms is also a linear equation. In this work we will treat the 
dominant $P_{33}$ channel. The model combines the diagrams of Fig.~\ref{xdir}b and 
Fig.~\ref{xdir}c, with unrenormalized couplings and masses, as the driving terms. We believe 
this to be the dominant physics in the $P_{33}$ channel. 

For the $P_{33}$ channel, the driving term for the combined model is
\begin{equation}
\langle\,q'\,\vert\,{\cal V}^{\, eff}(W)\,\vert\,q\,\rangle=
\lambda^{(0)}_N\,\frac{q'q\,v_{_N}(q')\, v_{_N}(q)}{-W_q+2m_{_N}-m^{(0)}_{_N}}+
\lambda^{(0)}_\Delta \,\frac{q'q\,v_{_\Delta}(q')\, 
v_{_\Delta}(q)}{W_q-m^{(0)}_{_\Delta}}\,\,\,, 
\label{cborn}
\end{equation}
where we have dropped the spin-isospin index $\alpha$ with the understanding that we 
are addressing specifically the $P_{33}$ channel. The algebra simplifies if we write 
the effective potential, Eq.~\ref{cborn}, as
\begin{equation}
\langle\,q'\,\vert\,{\cal V}^{\, eff}\,\vert\,q\,\rangle
=\sum_{i,j}\, v_i(q')q'\,\left(\sum_k G_{ik}^{(0)}(W)\lambda_{kj}^{(0)}\right)
v_j(q)q\,\,
\label{cbmatrix}
\end{equation}
with $i=1,2$ representing $N$ and $\Delta$ respectively, and
\begin{equation}
G_{ij}^{(0)}\Rightarrow\pmatrix{
   (-W+2m_{_N}-m_{_N}^{(0)})^{-1}&0\cr
   0&(W-m_\Delta^{(0)})^{-1}\cr}\,\,\,,
\label{gzero}
\end{equation}
and
\begin{equation}
\lambda_{ij}=\delta_{ij}\lambda_i^{(0)}\,\,\,.
\label{lmatrix}
\end{equation}
Defining the T-matrix as
\begin{equation}
\langle\,q'\,\vert\,t(W)\,\vert\,q\,\rangle=
\sum_{i,j}v_i(q')q'\,\tau_{ij}(W)\,v_j(q)q\,\,\,,
\label{tmatrix}
\end{equation}
and inserting this and Eq.~\ref{cbmatrix} into the Lippman-Schwinger equation, 
Eq.~\ref{LpSw}, gives a $2\times 2$ matrix equation,
\begin{equation}
\left\{[G^{(0)}(W)]^{-1}-\lambda\,\varepsilon (W)\right\}\,\tau (W)=\lambda\,\,\,
\label{LSmatrix}
\end{equation}
where inverting $G^{(0)}(W)$ is trivial since it is diagonal, and $\varepsilon (W)$ is 
defined by
\begin{equation}
\varepsilon_{ij}(W)\equiv \int\frac {q''^2\,dq''}{4\omega_{q''}E_{q''}}\,\frac{
   q''^2v_i(q'')\,v_j(q'')}{W-W_{q''}+i\eta}\,\,\,.
\label{epsilon}
\end{equation}
The matrix $\tau_{ij}(W)$ is given explicitly by
\begin{eqnarray}
   &&\tau_{ij}(W)\Rightarrow \nonumber\\
   &&\pmatrix{\lambda^{(0)}_{_N}[W-m^{(0)}_\Delta-\lambda^{(0)}_\Delta
       \varepsilon_{\Delta\Delta}(W)]& 
       \lambda^{(0)}_\Delta \lambda^{(0)}_{_N}\varepsilon_{\Delta N}(W)\cr
       \lambda^{(0)}_{_N}\lambda^{(0)}_\Delta\varepsilon_{N \Delta}(W)&
\lambda^{(0)}_\Delta [-W+2m_{_N}-m_{_N}^{(0)}- 
\lambda^{(0)}_{_N}\varepsilon_{_{NN}}(W)]\cr}
 /{\cal D}(W),
 \label{cnom}
 \end{eqnarray}
 with
 \begin{eqnarray}
 {\cal 
D}(W)&=&\left(W-m_\Delta^{(0)}-\lambda_\Delta^{(0)}\varepsilon_{\Delta\Delta}(W)
             \right)\,
             \left(-W+2m_{_N}-m_{_N}^{(0)}-\lambda_{_N}^{(0)}\varepsilon_{_{NN}}(W)
             \right)\nonumber \\
             &-&\lambda_\Delta^{(0)}\,\lambda_{_N}^{(0)}\varepsilon_{N\Delta}(W)\,
             \varepsilon_{\Delta N}(W)\,\,\,.
\label{determ}
\end{eqnarray}

We may remove the unrenormalized nucleon mass as a parameter by fixing the location of 
the nucleon pole in the scattering amplitude at its physical value. The 
nucleon pole occurs 
when ${\cal D}(W=m_{_N})=0$, which gives
\begin{equation}
m_{_N}^{(0)}=m_{_N}-\lambda_{_N}^{(0)}\varepsilon_{_{NN}}(m_{_N})
-\frac
{\lambda^{(0)}_{_N}\lambda^{(0)}_\Delta/,\varepsilon_{N\Delta}\varepsilon_{\Delta N}}
{m_{_N}-m_\Delta^{(0)}-\lambda^{(0)}_\Delta\,\varepsilon_{\Delta\Delta}(m_{_N})}\,\,\,
.
\label{crenorm}
\end{equation}
Algebraically eliminating the unrenormalized nucleon mass by substituting 
Eq.~\ref{crenorm} into Eqs.~\ref{cnom} and \ref{determ} does not yield any 
simplification. We thus adopt the numerical approach of using Eq.~\ref{crenorm} to 
calculate numerically the value of $m_{_N}^{(0)}$ and then use this value in 
calculating Eqs.~\ref{cnom} and \ref{determ}. We also do not find any simple expression 
for the renormalized pion-nucleon coupling constant. Rather than using complicated 
algebraic expressions, we calculate the renormalized coupling constant numerically by 
calculating the scattering amplitude near the nucleon pole. 

In Refs.~\cite{CBM,RJM84} approximate expressions for the scattering amplitude arising 
from the same Hamiltonian as is being used here were derived. In Ref.~\cite{CBM}, the 
Chew
series \cite{chew} was summed  approximately, while in Ref.~\cite{RJM84} a matrix 
$N/D$
approach was adopted. If we  ignore the coupling to inelastic channels (set $\hat
\eta=1$ in Ref.~\cite{RJM84}), these two  approaches produced
identically the same answer, something that seems to have been overlooked  probably
because of typos in both manuscripts. The question is how does this earlier  result
differ from that found here? The approximate summation of the Chew series in 
Ref.~\cite{CBM} is equivalent to the use here of the approximate crossing relation 
given in Eq.~\ref{cross}. Although the derivations are different, both produce the 
same approximation of the $U$-channel singularity as an $S$-channel singularity, so 
this is not a source of the resulting differences.

However, there  are two differences between the earlier works and our result. The 
first
is simply  kinematic. The invariant phase space used here produces a factor $2E_{q''}$
in the intermediate integration that is absent in the earlier work. Since  the earlier
works treat the form factor $v_{_N}(q)$ phenomenologically and adjust it to  fit data,
the form factor in these works contains implicitly this extra factor. This  is true of
a number of early models \cite{chlw}. Not explicitly including this phase space  
factor 
means
that it is implicitly included in the definition of $v_{_N}(q)$. The  range parameter
associated with $v_{_N}(q)$ would then necessarily be constrained to be near the nucleon
mass, or  approximately 1 GeV/c. 

In addition, the earlier models treat the renormalization of the nucleon mass 
differently than is done here. In the previous models, the renormalization of the 
nucleon mass would be given by Eq.~\ref{renormm}; the last term in Eq.~\ref{crenorm} 
would be absent. Renormalization is most easily understood in the absence of crossing. 
Think of a model for the $P_{11}$ channel with a direct nucleon pole and a Roper 
resonance, the $N^*{\rm (1440)}$. The physical nucleon would be a linear combination of 
the bare 
nucleon, the bare Roper, the bare nucleon plus a pion cloud, and the Roper plus a pion 
cloud. The mass renormalization would necessarily depend on the coupling constant 
$\lambda_{_{N^*}}^{(0)}$ and the form factor $v_{_{N^*}}(q)$. The residue of the
nucleon  pole must also contain terms with $v_{_{N^*}}(q)$ to reflect that the 
physical
nucleon  wave function contains an admixture of $N^*$. Since Eq.~\ref{renormm} is
independent  of $\lambda_{_{N^*}}^{(0)}$ and $v_{_{N^*}}(q)$, it cannot be a complete
and correct  description of the mass renormalization. The $P_{33}$ channel is more
subtle. In a  complete model, the crossed nucleon pole term must have a physical
nucleon with a mass  renormalization that is identical to the renormalization in the
direct nucleon pole  term. It is through the crossed nucleon pole term that the
$\Delta$ resonance enters the  mass renormalization. The underlying physics is that 
the nucleon contains a pion cloud plus bare delta coupled to $j=1/2$ component. The
additional terms included in the mass  renormalization in this work produce a more
physical, more complete, and more complex  model of the nucleon. However, as can 
easily be seen \cite{RJM82} in the simple  Chew-Low model, the renormalization of the 
nucleon mass and coupling constant will only  be independent of the spin-isospin 
channel if  the model is fully crossing symmetric. Thus  a definitive understanding of 
mass renormalization awaits the construction of such a  model. 

\section{Results}
\label{results}

These results, Eqs.~\ref{tmatrix}, \ref{cnom}, \ref{determ}, and \ref{crenorm}, are 
applied to elastic pion-nucleon 
scattering in the dominant $P_{33}$ channel. First, an extension of the model is to be 
made. For separable potential models \cite{pot}, the Chew-Low model 
\cite{chlw,DJE78b,DJE90}, and the Lee model \cite{DJE90}, the coupling of the pion-nucleon 
channel to inelastic 
meson-production channels was found to be significant. In both cases, $N/D$ arguments 
were used to incorporate into the model the effect of this coupling without having to 
model explicitly the inelastic channels. Since our model is equivalent to an 
energy-dependent potential model, the arguments from the original work \cite{pot} 
apply. The on-shell t-matrix in channel $\alpha$ is parameterized as
\begin{equation}
\langle\,q\,\vert\,t_\alpha(W_q)\,\vert\,q\,\rangle=
-\frac{4\hbar^2 W_q}{\pi q}\, \eta_\alpha\sin \delta_\alpha\,e^{i\delta_\alpha}\,\,\,.
\label{tparam}
\end{equation}
To include the effects of coupling to inelastic channels, the integral 
$\varepsilon_{ij}(W)$ in Eq.~\ref{epsilon} is to be replaced by
\begin{equation}
\varepsilon_{ij}(W)\equiv \int\frac 
{q''^2\,dq''}{4\omega_{q''}E_{q''}}\,\frac{1}{\eta(q'')}\,\frac{
   q''^2v_i(q'')\,v_j(q'')}{W-W_{q''}+i\eta}\,\,\,.
\label{epsilonp}
\end{equation}
The change is the inclusion in the integral of $\eta(q)^{-1}$, where $\eta(q)$ is defined 
by
\begin{equation}
\eta(q)\equiv \frac{\sigma_{in}(q)}{\sigma_{tot}(q)}\,\,,
\label{eta}
\end{equation}
with $\sigma_{in}(q)$ ($\sigma_{tot}(q)$) the measured inelastic cross section (total 
cross 
section) in channel $\alpha$.
The most general form of a potential which leads to this result is given in \cite{pot} 
while for the Lee model, this form results \cite{DJE90} from the doorway concept --- 
the system couples only to the inelastic channels by first proceeding through a 
resonant state. Unitarity in the presence of inelastic channels as embodied in 
Eq.~\ref{tparam} is identically satisfied by the use of Eq.~\ref{epsilonp}.

We assume that the form factor for coupling to the nucleon and to the $\Delta$ are 
identical. We choose
\begin{equation}
v_{_N}(q)=v_\Delta(q)= v(q)=e^{-q^2/\beta^2}\,\,.
\label{ff}
\end{equation}
The identity of these form factors follows from the assumption that the bare 
nucleon and the bare $\Delta$ are composed of valence quarks with the same spatial 
structure, differing only in their spin-isospin structure. The selection of a Gaussian 
as the functional form could be motivated by a constituent quark model \cite{Isgur}.
However, previous work has indicated little sensitivity to the specific function 
chosen for the form factor. It is best to view this simply as a choice of a convenient 
function that provides a cutoff with a range parameter to be determined by the data.

\begin{figure}
\epsfxsize=6.5in
\epsfbox{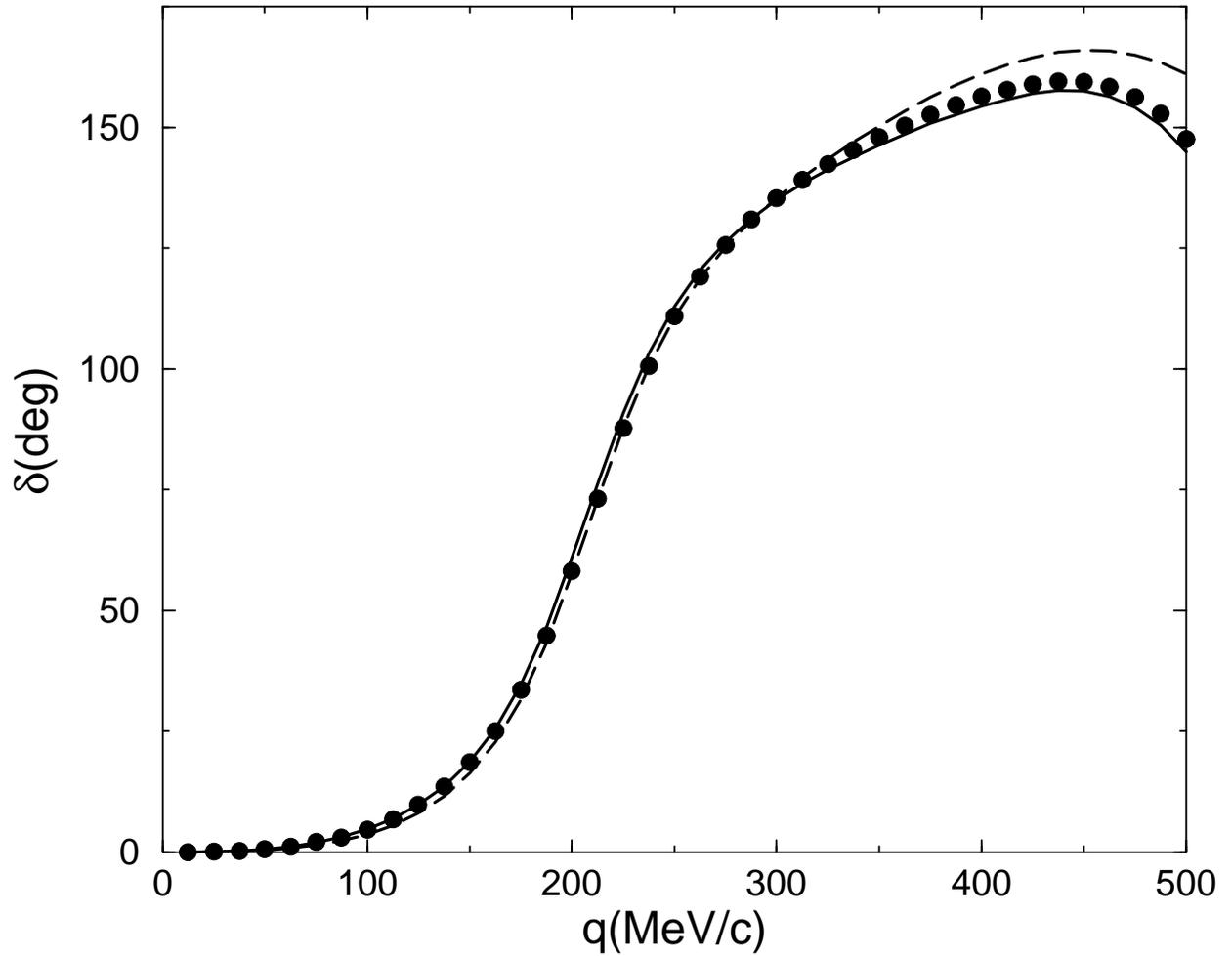}
\vspace {25pt}
\caption
{The phase shift $\delta_{33}$ in the $P_{33}$ channel versus the center-of-momentum 
momentum $q$. The dots are the result of the phase shift analysis of 
Ref.~\protect\cite{SAID}. The solid curve is the results of the Chew-Low model and the 
dashed curve is the results for the Lee model.}
\label{fig3}
\end{figure}

Before examining the combined model, we first examine results from the Chew-Low model and 
the 
Lee 
model separately.  This will help us to understand the results that emerge from the 
combined model. Once the coupling to the inelastic channels has been incorporated into 
the Chew-Low model, it produces results \cite{chlw,DJE78b,DJE90} which are an excellent 
reproduction of the data. We depict this in Fig.~\ref{fig3} where we plot the phase of 
the scattering amplitude $\delta_{33}(q)$, Eq.~\ref{tparam}, versus the center-of-momentum 
momentum $q$. The dots are the data from Ref.~\cite{SAID} and the solid curve 
is the result of the Chew-Low model. This two parameter, a coupling constant and 
a range for the form factor, model not only fits well the region dominated by the 
$\Delta$, 
$q\le$ 300 MeV/c but continues to fit well for several hundred MeV above this $\Delta$ 
region. The data in the region from threshold to $q=$ 300 MeV/c is determined by three 
parameters --- the position and the width of the $\Delta(1232)$ and the behavior of 
the 
phase $\delta_{33}$ as it approaches zero. The Chew-Low model, generalized to include 
the coupling to inelastic channels, naturally reproduces with two parameters the three 
parameters which characterize the data.

The difficulty with the  Chew-Low model is that it does not contain a quark $\Delta$ 
state and the excellent fit results \cite{DJE90} from a cutoff given by $\beta=$ 2285 
MeV/c. This is a much higher momentum cutoff than is indicated by any other data.
Earlier \cite{chlw} applications of the Chew-Low model did not include the nucleon 
phase space factor and thus they gave $\beta\sim$ 1 GeV, but this was because the 
nucleon phase space had been implicitly contained in the definition of the form factor 
in these works.

\begin{figure}
\epsfxsize=6.5in
\epsfbox{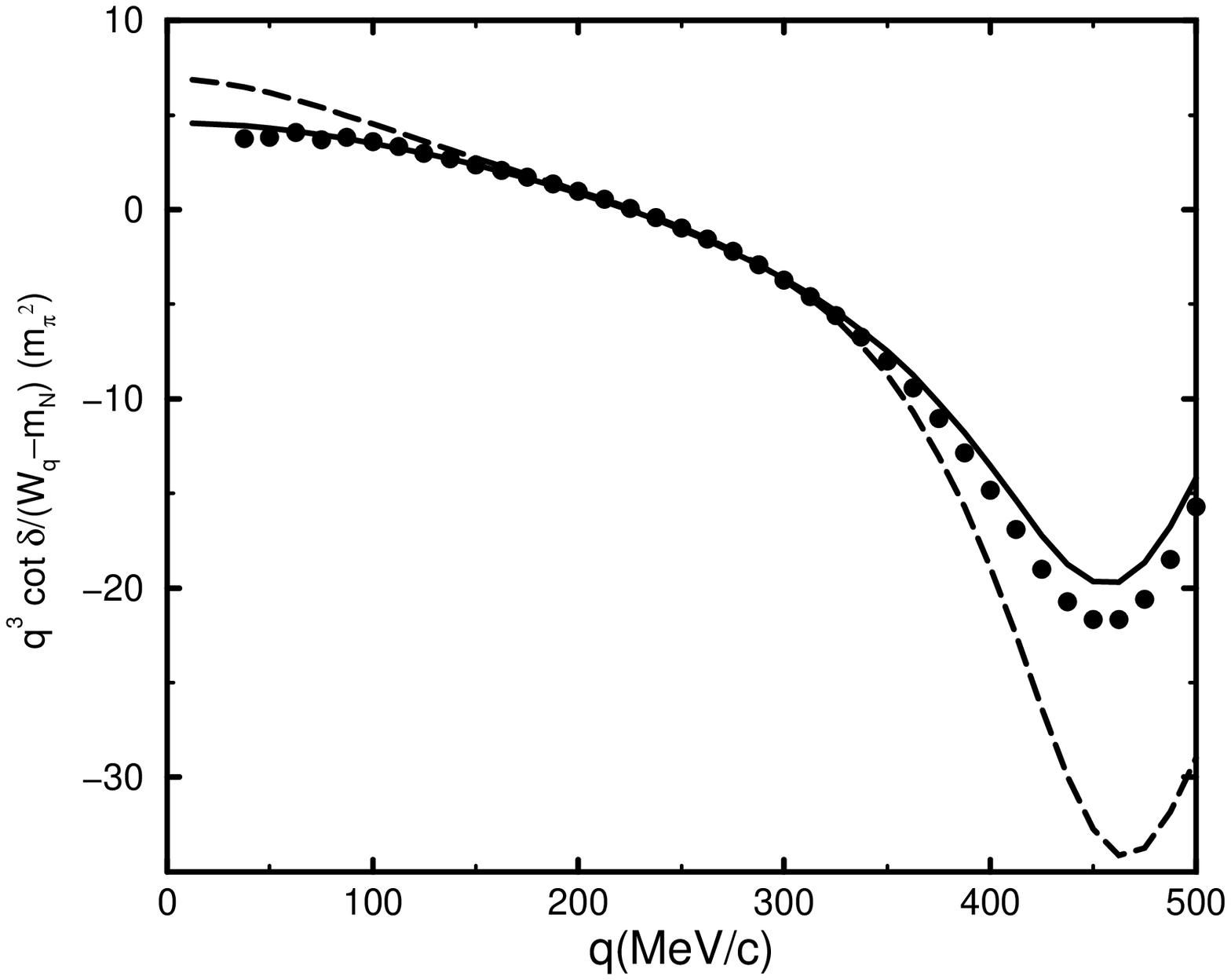}
\vspace {25pt}
\caption
{The same as Fig.~\protect\ref{fig3} except the quantity $q^3\,\cot 
\delta/(W_q-m_{_N})$ is presented.}
\label{fig4}
\end{figure}

The Lee model alone is not expected to fit well the data. This is because the 
low-energy data is dominated by the nucleon pole and the scattering amplitude from 
this 
model does not contain this pole. It has been pointed out \cite{DJE78b} that the data 
can indeed be 
fit but that this requires a factor of $\omega_q^{-1/2}$ in the form factor, i.e. an 
artificially low momentum cutoff. The best fit for the Lee model is shown as the 
dashed 
line in Fig.~\ref{fig3}. In order to better understand this result, we plot in 
Fig.~\ref{fig4} the quantity $q^3\,\cot \delta/(W_q-m_{_N})$. This quantity removes 
the 
$q^3$ threshold behavior and also removes the energy dependence $(W_q-m_{_N})^{-1}$ 
induced by the nucleon 
pole. The solid curve in Fig.~\ref{fig4} is again the Chew-Low curve. This curves 
demonstrates better the quality of the fit for $q\le $ 300 MeV, and emphasizes more 
the 
difference between the data and the model at the higher energies. The dashed curve in 
Fig.~\ref{fig4} is the best fit results for the Lee model. This demonstrates that 
this 
model is able to fit the position and the width of the $\Delta$ but not the data below 
and above the resonance. The fit presented here is a compromise at fitting reasonably 
the data both below and above the resonance. One can fit well the data below the 
resonance, for example, but then the fit just above the resonance, $q\geq$ 200 MeV 
becomes very poor. This is even though the model has three free parameters --- the 
coupling constant, the form factor cutoff range, and the bare mass of the $\Delta$. 
The 
range of the form factor for the fit presented is $\beta=$ 400 MeV/c. 

\begin{figure}
\epsfxsize=6.5in
\epsfbox{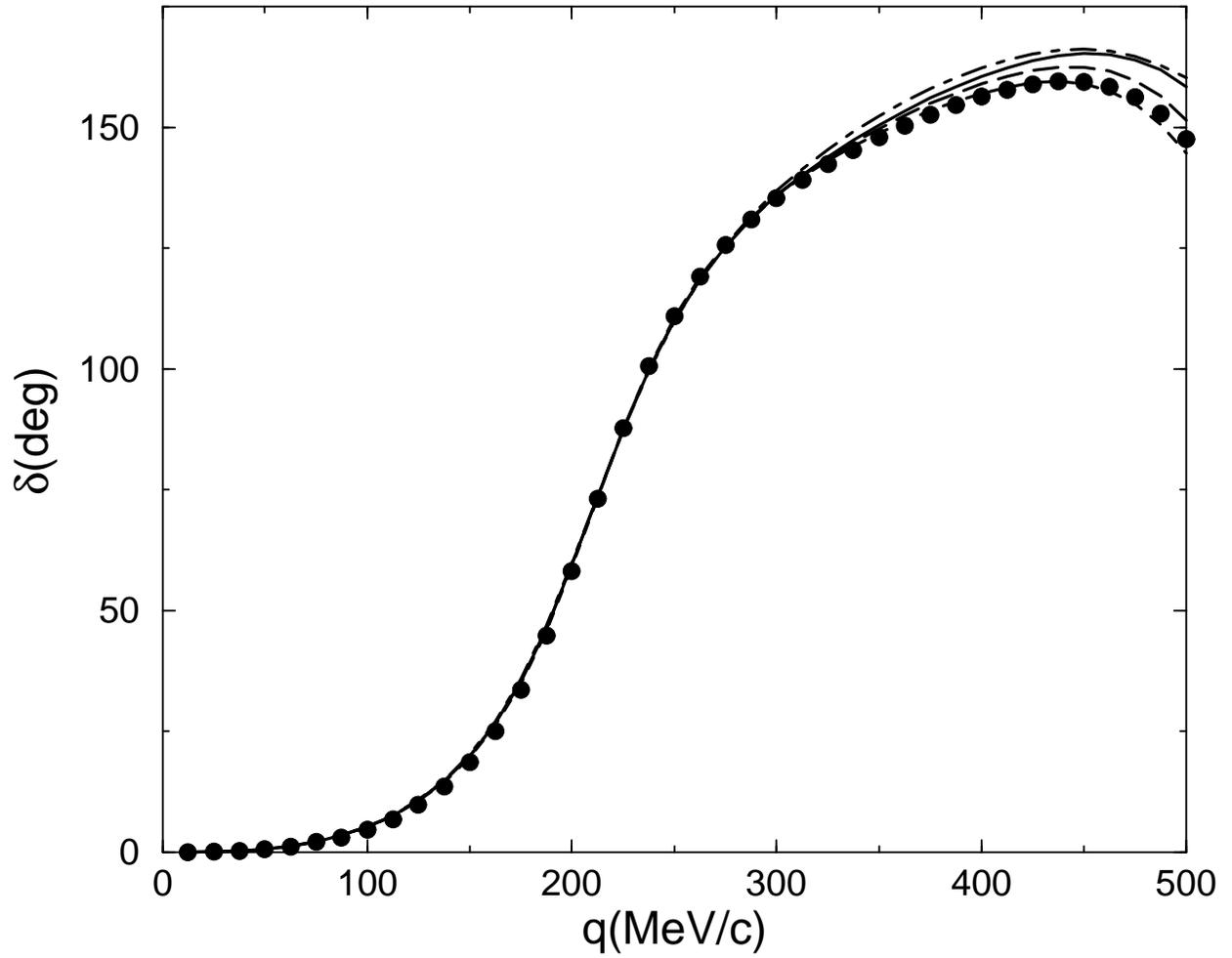}
\vspace {25pt}
\caption
{The same as Fig.~\protect\ref{fig3} except the curves are the results of the combined 
model. The solid curve corresponds to $\beta=$ 1100 MeV/c, the long-dashed curve to 
$\beta=$ 800 MeV/c, the short-dashed curve to $\beta=$ 500 MeV/c, and the dot-dashed 
curve to $\beta=$ 400 MeV/c.}
\label{fig5}
\end{figure}

The question that these results present is how can a model which combines the two 
interactions,  $\pi + N\leftrightarrow N,\,\Delta(1232)$, be accommodated by 
the 
data? The answer is given in Fig.~\ref{fig5} where we present four curves which are 
all 
reasonable fits to the data. The curves correspond to four values of the cutoff 
parameter, $\beta=$ 400, 500, 800, and 1100 MeV/c. These are four values from the 
continuum 
set of values of $\beta$ which produce good fits to the data. The values for the 
parameters of 
the model that correspond to these values of $\beta$ are given in Table~\ref{table1}.

\begin{figure}
\epsfxsize=6.5in
\epsfbox{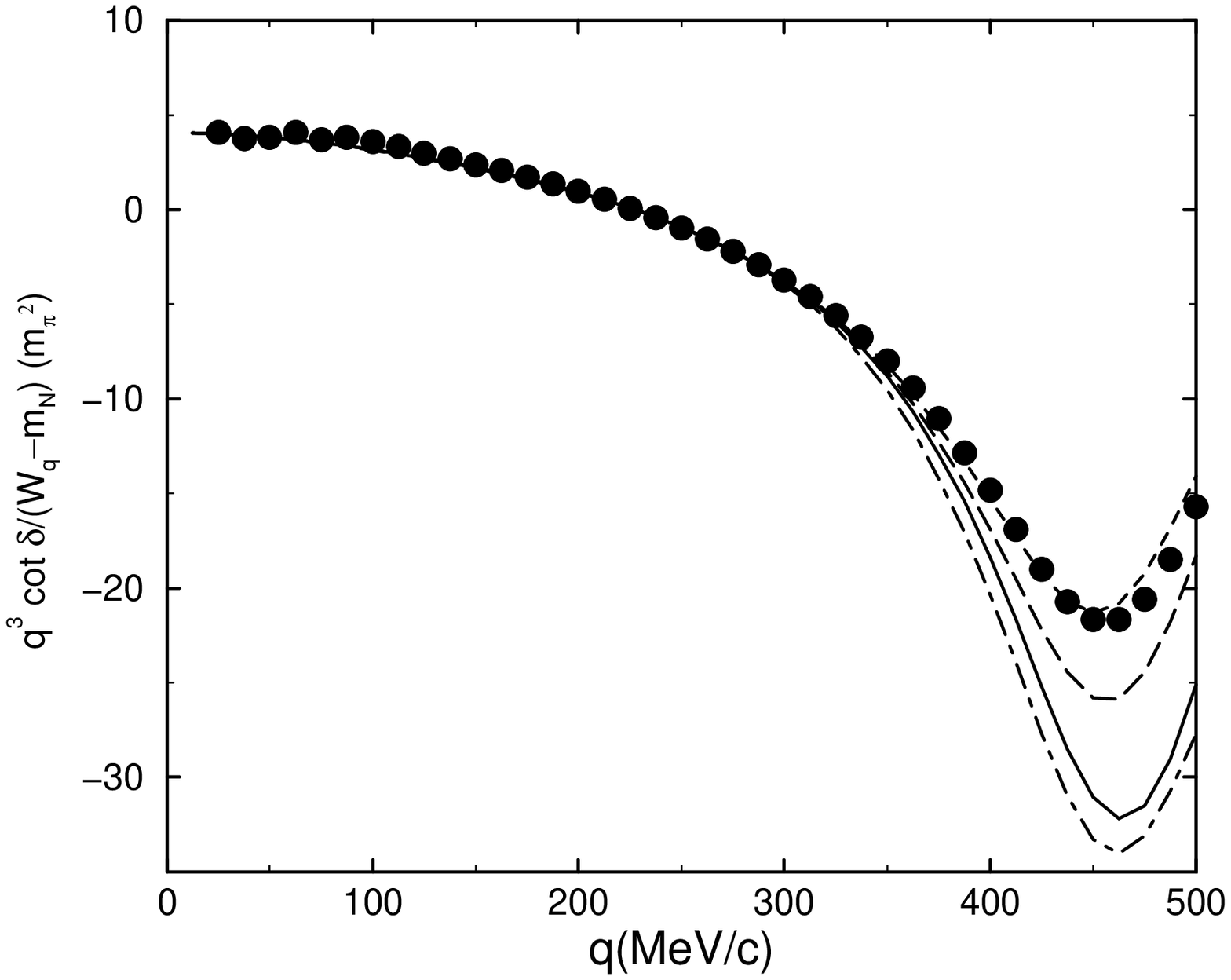}
\vspace {25pt}
\caption
{The same as Fig.~\protect\ref{fig5} except the quantity $q^3\,\cot 
\delta/(W_q-m_{_N})$ is presented.}
\label{fig6}
\end{figure}

As the Chew-Low model 
already reproduces well the data, we find a continuum of solutions for the combined 
model. The combined model contains four free parameters, the range of the form factor, 
two coupling constants, and the bare mass of the $\Delta$. The data is able to fix 
three out of the four parameters, but not all four. In Fig.~\ref{fig6} we again 
present 
the quantity $q^3\,\cot \delta/(W_q-m_{_N})$. We see that the fits are excellent for 
$q\le $ 300 
MeV/c. Above this region, we do not require an exact fit to the phase shifts. 
Comparing 
Figs.~\ref{fig4} and \ref{fig6} we see that the curves for the combined model with 
$\beta=$ 400 MeV/c and $\beta=$ 1100 MeV/c are inferior to the Chew-Low model. In this 
case we have found a local minimum as the true minimum would be to set the $\Delta$ 
coupling to zero and use the Chew-Low results.
 
\begin{figure}
\epsfxsize=6.5in
\epsfbox{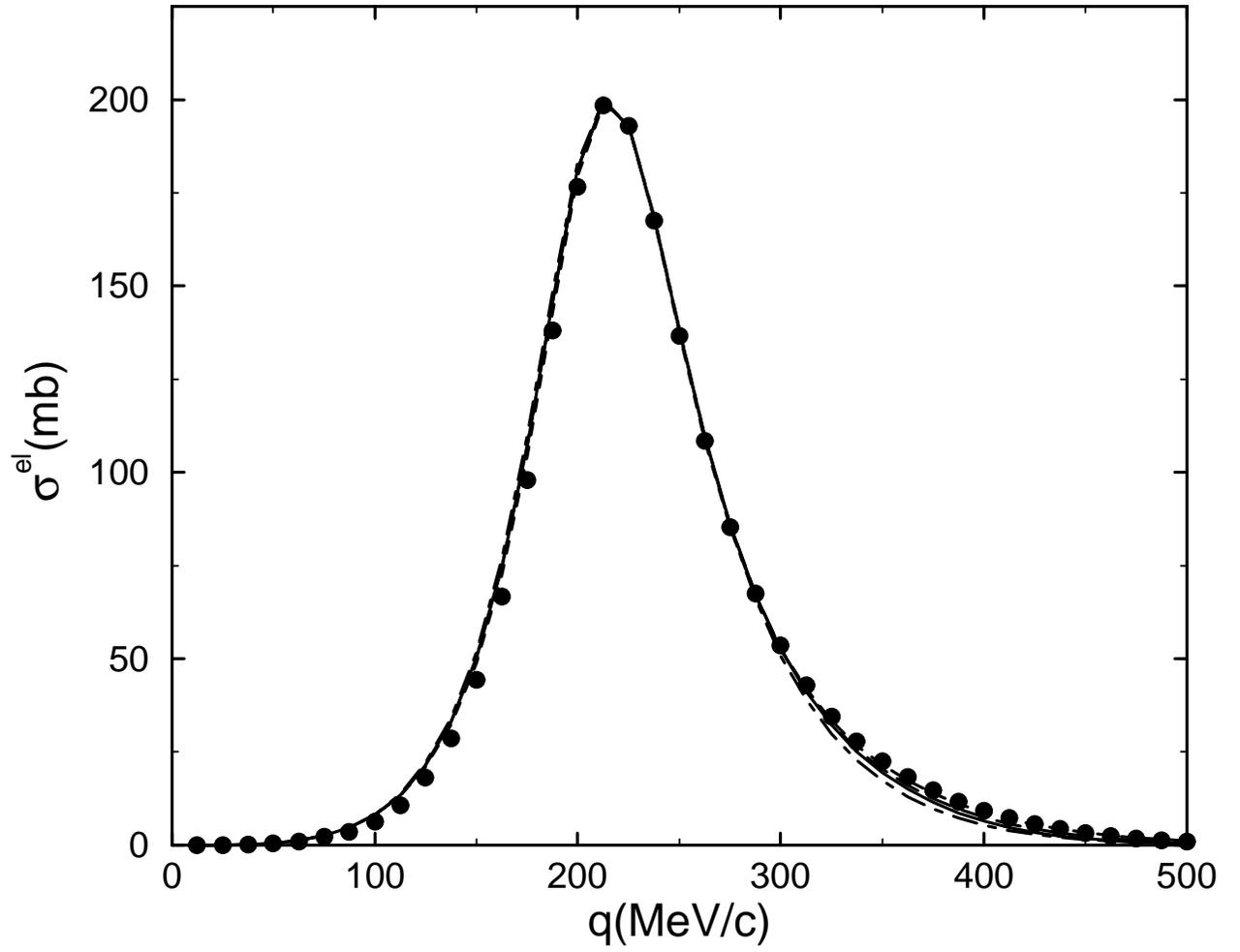}
\vspace {25pt}
\caption
{The same as Fig.~\protect\ref{fig5} except the total elastic cross section, 
$\sigma_{e\ell}^{tot}$, in the $P_{33}$ channel is presented. }
\label{fig7}
\end{figure}

We believe Fig.~\ref{fig6} to be somewhat misleading. Above the the resonance, the 
$P_{33}$ amplitude is quite small and does not contribute significantly to 
pion-nucleon 
scattering. This is illustrated in Fig.~\ref{fig7} where we plot the total elastic 
cross section, $\sigma_{e\ell}^{tot}$. The four curves for the four values of $\beta$ 
are plotted, but because they differ only by an amount that is about a line width, they 
are hard to distinguish. The lower limit on $\beta$ of 400 MeV/c is firm. 
Going lower than this gives results which are not compatible with the data for $q\le$ 
300 MeV/c. Our choice of an upper limit of 1100 MeV/c is not so firm. If we were to 
include only data below 300 MeV/c then excellent fits would result for $\beta$ 
extending all the way up to the Chew-Low results of 2250 MeV/c. The upper limit of 1100 
MeV/c results from requiring a fit  in the region of $q\approx$ 500 MeV/c.

We are fitting phase shifts which are not data themselves, but 
parameters extracted from data. This prohibits a statistical analysis of what is an 
acceptable fit. 
However, the results given in Ref.~\cite{SAID} indicate that the phases above $q=$ 300 
MeV/c are well determined so we include a criteria of a {\it reasonable} fit to these 
data, where we define {\it reasonable} by making a judgment from the results in 
Figs.~\ref{fig5} and \ref{fig6}. Allowing $\beta$ to be larger than 1100 MeV/c gives 
curves which are significantly further away from the data in the region $q\sim$
500 MeV/c. 

Another consideration is that there are theoretical systematic errors. The assumption we 
have 
made for the underlying interaction does not include a small four-point interaction 
which might be important for $q\geq$ 400 MeV/c. We have assumed an infinite nucleon 
mass form for the crossed driving terms; there might be small corrections to this in 
this region. We have used the no crossing approximation assuming that 
increasing the residue of the nucleon pole term would compensate. This is true over a 
limited momentum region, and we do not know how accurately and over what region this is 
valid. Thus a value for $\beta$ greater than 1100 MeV/c cannot be absolutely excluded.

What is certain is that the data in the $P_{33}$ channel is not sufficient to uniquely 
determine the parameters of the model. This data will fix three of the parameters as a 
function of a fourth. We find, if we impose a fit to the phase shifts in the region 
near $q=$ 400 MeV/c, $\beta =750 \pm 350$ MeV/c. The same criteria would also allow the 
Chew-Low model as a satisfactory fit to the data. The values of the unrenormalized 
coupling constants, $\lambda_{_N}^{(0)}$ and $\lambda_\Delta^{(0)}$, are depicted in 
Fig.~\ref{fig7} as a function of the cutoff parameter $\beta$. We see that for the 
larger values of $\beta$ the theory is fitting the data with a model that is primarily 
the Chew-Low model; the small differences between the Chew-Low model and the data is 
being corrected by a small addition of the coupling to the $\Delta$. As the cutoff 
$\beta$ decreases, the balance shifts. At the lowest value of $\beta$, 400 MeV/c, the 
interaction is predominantly the coupling to the $\Delta$ but with a not negligible
contribution from the Chew-Low interaction.  In Fig.~\ref{fig7} we also depict the 
renormalized pion-nucleon coupling constant as a function of $\beta$. For $\beta$ greater 
than about 500 MeV/c, the renormalized coupling constant is reasonably independent of 
$\beta$. The renormalized coupling constant obtains from an extrapolation of the low 
energy 
data to the subthreshold energy $W_q=m_{_N}$ and thus should be approximately independent 
of 
the model. We find for $f_{\pi NN}^2$ the range $f_{\pi NN}^2=0.142\pm .004$ if we restrict 
the range of $\beta$ to 500 to 1100 MeV/c. This is larger than the value \cite{RA95}, 
$f_{\pi N}^2=0.076$, recently extracted from nucleon-nucleon scattering. The difference 
arises, as mentioned earlier, because we have neglected the left-hand crossing cut 
depicted in Fig.~\ref{anal} and compensated by an increase in the coupling constant.

The renormalization constant $Z_c\equiv \lambda_{_N}/\lambda_{_N}^{(0)}$ gives an 
indication of whether the mesonic cloud effects can be treated
perturbatively. We find $Z_c=$ 1.25 for $\beta=$ 400 MeV/c and 1.51 for $\beta=$ 500 
MeV/c. From there it rises rapidly to a value of 2.46 for $\beta=$ 1100 MeV/c. Thus a 
perturbative treatment of the mesonic cloud is not adequate except in the region of 
low cutoffs below about 500 MeV/c.

Invoking SU(6) would fix the ratio of the coupling 
constants
\begin{equation}
R=\frac{\lambda_\Delta^{(0)}}{\lambda_{_N}^{(0)}}=\left(\frac{f_{\Delta N\pi}^{(0)}}
{f_{NN\pi}^{(0)}}\right)^2\,\,.
\label{ccrel}
\end{equation}
This would provide one additional relationship among the parameters and give a unique 
solution for the model, as was done in \cite{CBM}. However, none of the solutions which we find 
has a value 
of $R$ as large as the SU(6) prediction.

Of the continuum of solutions which we find, those with 
smaller $\beta$ are of the same character as the solution proposed in Ref.~\cite{CBM}.
These solutions have a relatively low momenta cutoff and describe the physical resonance as 
predominantly arising from the $\Delta$ with small corrections from the Chew-Low 
interaction. 

\begin{figure}
\epsfxsize=6.5in
\epsfbox{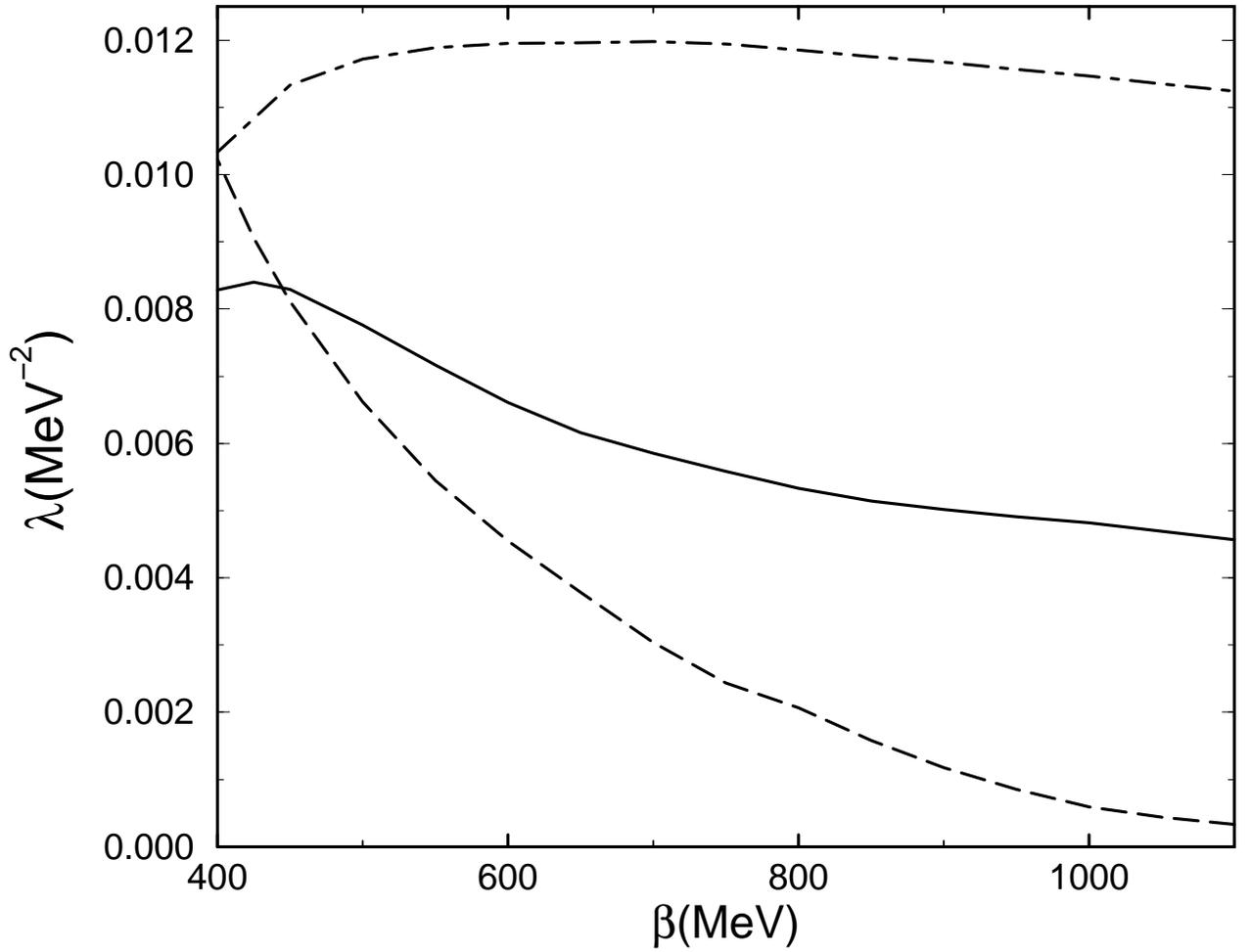}
\vspace {25pt}
\caption
{The coupling constants as a function of the form-factor cutoff parameter $\beta$ for 
values which fit the $P_{33}$ data. The 
solid curve is the unrenormalized nucleon coupling $\lambda_{_N}^{(0)}$, the dashed 
curve is the unrenormalized $\Delta$ coupling constant $\lambda_\Delta^{(0)}$, and 
the 
dot-dashed curve is the renormalized nucleon coupling constant $\lambda_{_N}$. }
\label{fig8}
\end{figure}

The model developed here allows one to extract the bare mass of the nucleon and the  
the $\Delta$. The mass of these baryons in the absence of the 
coupling to mesons can be associated with the mass of the the state made up only of 
valence 
quarks. Symmetry arguments should be more valid for the simple valence quark 
states than for the more complex physical particles. In Fig.~\ref{fig8} we present the 
bare mass of the nucleon and the $\Delta$ as a function of the cutoff $\beta$. The 
bare mass of the $\Delta$, $m_\Delta^{(0)}$, is one of the parameters fit to the data. 
The 
bare mass of the nucleon, $m_{_N}^{(0)}$, is calculated from Eq.~\ref{crenorm}. As 
$\beta\rightarrow$ 0, the bare masses approach the physical masses. The nucleon bare 
mass rises nearly linearly with $\beta$ reaching a value of about 1300 MeV for $\beta=$ 
1100 MeV/c. On the other hand, the $\Delta$ bare mass rises to a maximum of 1700 MeV 
for $\beta$ near 850 MeV/c and then falls slowly. For $\beta=$ 1300 MeV/c the curves 
cross 
and the bare $\Delta$ mass becomes smaller than the bare mass of the nucleon. 

An important number is the difference in the bare masses, $\delta m^{(0)} = 
m_{_N}^{(0)}-m_\Delta^{(0)}$. In a quark model this difference would be accounted for 
by a residual gluon exchange interaction. We find $\delta m^{(0)}=$ 330 MeV for $\beta 
=$ 400 MeV/c, as compared to 294 MeV, the difference between the energy at which the 
$\Delta$ 
resonance occurs and the nucleon mass. The mass difference reaches a 
peak value of 450 MeV for $\beta =$ 850 MeV/c and falls to 225 MeV for $\beta=$ 1100 
MeV/c.

For the Chew-Low model, the incorporation of the coupling to the inelastic channels 
\cite{chlw} enabled the model to fit well the data. We find that setting $\eta(q)$ equal 
to 
one in Eq.~\ref{epsilonp}, thus neglecting the coupling to inelastic channels, does not 
prevent excellent fits to the data. Although no longer necessary for a good fit, the 
coupling to the inelastic channels is a real physical phenomenon and thus including $\eta(q)$ 
is 
the more physical model. The inclusion of $\eta(q)$ generalizes the model effectively to 
include the coupling of the nucleon and $\Delta$ to any meson-baryon or multi-meson baryon 
channels without having to model those channels explicitly.

\section {Conclusions}

We have examined the question of how to solve for the elastic scattering amplitude 
when 
the underlying Hamiltonian is assumed to be of the form $\pi + N\leftrightarrow 
N,\,\Delta(1232)$. We provide a new solution that is 
an extension of the work in Refs.~\cite{CBM} and \cite{RJM84}. The model makes use of 
the observation that the Chew-Low model in the no crossing approximation, with either a 
direct or crossed driving term, is a 
linear model when written in terms of the  unrenormalized mass and coupling constant. 
The new model, although quite similar to the earlier models, differs in the way that 
it treats the renormalization of the nucleon mass.

The phase shifts in the dominant $P_{33}$ channel were fit by the model. However, the 
data are not capable of uniquely determining the parameters of the model. Good fits to 
the data are found for a continuum of values for the model parameters. We find the 
cutoff range for the pion-nucleon form factor to be given by $\beta = 750\pm350$ 
MeV/c.
Perturbative treatments of the mesonic cloud are found not to be accurate unless the 
cutoff parameter is in the low range, $\beta\leq$ 500 MeV/c. Of the continuum of solutions 
found, a subset with a low momentum cutoff and a $\Delta$ 
resonance that is predominantly the bare $\Delta$ is qualitatively similar to the Cloudy Bag 
solution \cite{CBM}.
An important feature of the model is its ability to calculate the unrenormalized 
masses 
of the nucleon and the $\Delta$. For the nucleon, we find $m_{_N}^{(0)}= 1200\pm 200$ 
MeV, and for the $\Delta$, $m_\Delta^{(0)} = 1500\pm 200$ MeV. The difference in the 
bare masses, a quantity which would be accounted for by a residual gluon interaction, 
is found to be $\delta m^{(0)}=350\pm 100$ MeV.

Since the $P_{33}$ data alone are not capable of uniquely determining the parameters 
of 
the model, a further generalization of the model is needed. If we are to use 
pion-nucleon scattering to determine the parameters of the model, then the next step 
would be to include additional spin-isospin channels. A crossing symmetric model would 
require the simultaneous treatment of the $P_{11}$, $P_{13}$, $P_{31}$, and $P_{33}$ 
channels. Several techniques have been developed \cite{RJM82} to solve the infinite 
nucleon mass, crossing symmetric Low equation for the Chew-Low interaction. The model 
used here is already more complex than the simple Chew-Low model and in order to 
produce physical results would have to be expanded to include the $\pi + 
N\leftrightarrow N^*{\rm (1440)}$ interaction. Ways of generalizing the formalism of 
\cite{RJM82} to this more complex situation are being investigated.

In Ref.~\cite{DJE90} the Lee model was used to fit the D- and F-wave pion-nucleon 
resonances. The unrenormalized resonant mass were observed \cite{DJE95} to be more nearly 
degenerate than the physical resonance energies. The model did not use form factors which 
were  consistent with each other. Each channel had a Gaussian form factor with a range that was 
independently adjusted. Crossing symmetry was also not included. The model was developed 
as input for pion-nucleus calculations \cite{MWRS} and not intended to address the 
question of the bare masses of the baryons. It will be interesting to see if a more 
consistent model produces bare masses which remain more nearly degenerate.

\acknowledgements
This work was supported in part by the US Department of Energy under grant 
DE-FG02-96ER40975.  The Southeastern Universities Research Association (SURA) operates 
the Thomas Jefferson National Accelerator Facility under DOE contract DE-AC05-84ER40150.

\vfill\eject 
\vfill\eject
\vfill\eject 
\vfill\eject 
\vfill\eject 
\vfill\eject
\vfill\eject 
\vfill\eject 
\begin{figure}
\epsfxsize=6.5in
\epsfbox{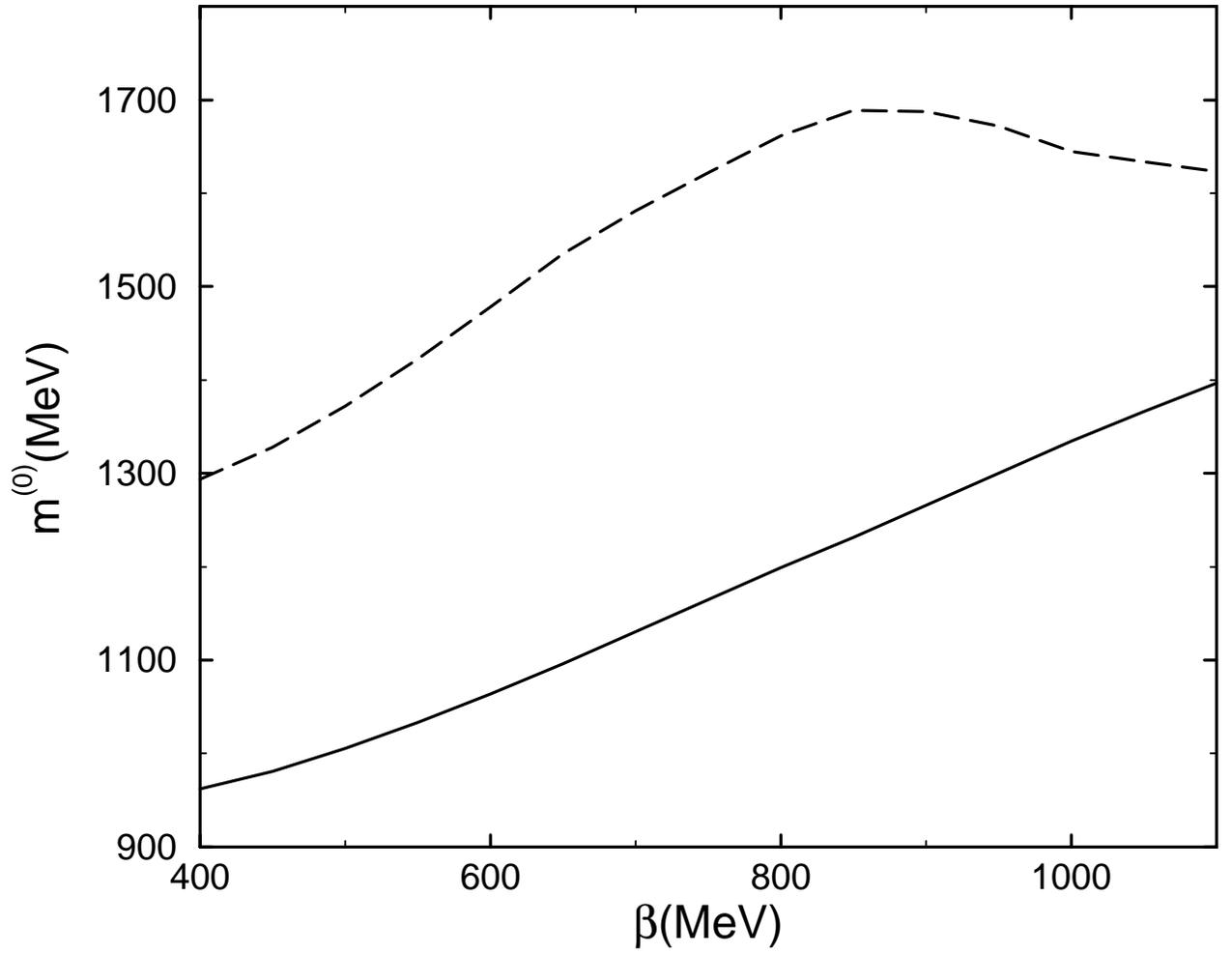}
\vspace {25pt}
\caption
{The bare masses, i.e. the masses in the absence of meson couplings, of the nucleon, 
solid curve, and the $\Delta$, dashed curve, as a function of the form-factor cutoff 
parameter $\beta$.}
\label{fig9}
\end{figure}
\vfill\eject 
\begin{table}
\caption{Typical sets of parameters (the form-factor cutoff $\beta$, the unrenormalized 
coupling constants $\lambda^{(0)}_i$, and unrenormalized $\Delta$ mass
$m_\Delta^{(0)}$) which produce fits to the $P_{33}$ phase shifts. Also given for each 
fit are two calculated parameters, $\lambda_{_N}$ the renormalized pion-nucleon 
coupling strength, and $m_{_N}^{(0)}$ the bare nucleon mass. These sets of parameters 
correspond to the curves depicted in Figs.~\protect\ref{fig5}--\protect\ref{fig7}. The numbers 
are given to an accuracy such that the results of this work can be reproduced.
\label{table1} }
\vskip 5pt
\begin{tabular}{cccccc}
\quad $\beta$ (MeV/c)\quad & \quad $\lambda_{_N}^{(0)}$ (MeV$^{-1}$) \quad 
& \quad $\lambda_\Delta^{(0)}$ (MeV$^{-1}$)  \quad  & 
\quad $\lambda_{_N}$ (MeV$^{-1}$)  & \quad $m_{_N}^{(0)}$ (MeV) 
 \quad  &\quad $m_\Delta^{(0)}$ (MeV)\\
\hline
400    & $8.28\times 10^{-3}$ & $1.02\times 10^{-2}$ & $1.03\times 10^{-3}$ & 962 & 
   1294 \\
500    & $7.77\times 10^{-3}$ & $6.62\times 10^{-3}$ & $1.17\times 10^{-3}$ & 1005 &
   1372 \\
800    & $5.34\times 10^{-3}$ & $1.98\times 10^{-3}$ & $1.19\times 10^{-3}$ & 1198 &
   1662 \\
1100   & $4.57\times 10^{-3}$ & $3.33\times 10^{-4}$ &$ 1.12\times 10^{-3}$ & 1397 &    
   1623 \\
\end{tabular}
\end{table}


\begin{references}

\bibitem{NIbook} see, for example, N. Isgur, {\it Quark Confinement and the Hadron     
    Spectrum III; Proceedings from the Institute for Nuclear Theory - Vol. 8,} (World  
    Scientific, Singapore, 1998);
    V. D. Burkert, L. Elouadrhiri. J. J. Kelley, and R. C. Minehart, {\it Excited 
Nucleons 
    and Hadronic Structure}, (World Scientific, Singapore, 2001).

\bibitem{Isgur} N. Isgur and G. Karl, Phys. Rev. D {\bf 18}, 4178 (1978);
    D {\bf 19}, 2653 (19879);
    D {\bf 23}, 817 (1981);
    S. Kapstick and N. Isgur, Phys. Rev. D {\bf 34}, 2809 (1986);
    S. Kapstick, Phys. Rev. D {\bf 46}, 1965 (1992); {\bf 46}, 2864 (1992).

\bibitem{LYG96} L. Ya. Glozman and D. O. Riska, Phys. Rep. {\bf  268}, 263 (1996).

\bibitem{MIT}A. Chodos, R. L. Jaffe, K. Johnson, C. B. Thorn, and V.
    F. Weisskopf, Phys. Rev. D {\bf 9}, 3471 (1974);
    T. DeGrand, R. L. Jaffe, K. Johnson, and J. Kiskis, Phys. Rev. D {\bf 12}, 2060 
    (1975).

\bibitem{CBM} G. A. Miller, A. W. Thomas, and S. Th\'eberge, Phys. Lett. {\bf 91B}, 192 
(1980); A. W. Thomas, S. Th\'eberge and G. A. Miller, 
    Phys. Rev. D  {\bf 22}, 2838 (1980); {\bf 23}, 2106E (1981); {\bf 24}, 216 (1981).

\bibitem{picloud}
    L. A. Copley, G. Karl. and E. Obryk, Nucl. Phys. {\bf B13}, 303 (1969);
    F. Foster and G. Hughes, Z. Phys. C {\bf 14}, 123 (1982);
    R. Koniuk and N. Isgur, Phys. Rev. Lett. {\bf 44}, 845 (1980);
    Phys. Rev. D{\bf 21}, 1868 (1980);
    Z. Li and F. E. Close, Phys. Rev. D {\bf 42}, 2207 (1990);
    S. Capstick and W. Roberts, Phys. Rev. D {\bf 47}, 1994 (1993);
    {\bf 49}, 4570 (1994); 
    {\bf 57}, 4301 (1998);
    R. Bijker, F. Iachello, and A. Leviatan, Ann. Phys. (NY) {\bf 236}, 69 (1994);
    Phys. Rev. D {\bf 55}, 2862 (1997);
    A. Chodos and C. B. Thorn, Phys. Rev. D {\bf 12},  2733 (1975). 

\bibitem{SAID}
    R. A. Arndt, Interactive Dial-In (SAID) Program, George 
    Washington University;
    R. A. Arndt, I. I. Strakowsky, and R. L. Workman, Phys. Rev. C {\bf 52}, 2875       
     (1995).

\bibitem{pot} D. J. Ernst, J. T. Londergan, E. J. Moniz and R. M. Thaler,  Phys.    
    Rev. C {\bf 10}, 1708 (1974); 
    J. T. Londergan, K. W. McVoy, and E. J. Moniz, Ann. Phys. (N.Y.) {\bf 86}, 147 
    (1974);
    C. Coronis and R. Landau, Phys. Rev. C {\bf 24}, 605 (1981);
    
\bibitem{LM85}L. Mattelitsch and H. Garcilazo, Phys. Rev. C {\bf 32}, 1635 (1985).

\bibitem{chlw} C. B. Dover, D. J. Ernst, R. A. Friedenberg, and R. M. Thaler,
    Phys. Rev. Lett {\bf 33}, 728 (1974);
    K. K. Kumar and Y. Nogami, Phys. Rev. D {\bf 20}, 2626 (1979); D {\bf 22}, 2098   
    (1980); 
    D. J. Ernst and M. B. Johnson, Phys. Rev. C {\bf 22}, 651 (1980).
    
\bibitem{DJE78b}    
    D. J. Ernst and M. B. Johnson,  Phys. Rev. C {\bf 17}, 247 (1978).

\bibitem{DJE90} D. J. Ernst, G. E. Parnell and C. Assad, Nucl. Phys. {\bf A518}, 
    658 (1990).

\bibitem{RJM81} R. J. McLeod and D. J. Ernst, Phys. Rev. C {\bf 23}, 1660 (1981).
    
\bibitem{JBC73} J. B. Cammarata and M. Banerjee, Phys. Rev. Lett. {\bf 31}, 610        
    (1973); Phys. Rev. C {\bf 13}, 299 (1976).

\bibitem{tshl} T. Yoshimoto, T. Sato, M. Arima, and T.-S. H. Lee, Phys. Rev.
    C {\bf 61}, 065203 (2000);
    R. J. McLeod and D. J. Ernst, Phys. Rev. C {\bf 49}, 1087 (1994);
    M. Arima, K. Shimizu, and K. Yazaki, Nucl. Phys. {\bf A543}, 613 (1992).

\bibitem{MGF95} M. G. Fuda, Phys. Rev. C {\bf 52}, 2875 (1995).

\bibitem{FG93} F. Gross and Y. Surya, Phys. Rev. C {\bf 47}, 703 (1993).

\bibitem{BCP91} B. C. Pearce and B. K. Jennings, Nucl. Phys. {\bf A528}, 655 (1991).

\bibitem{CS94} C. Sch\"utze, J. W. Durso, K. Holinde, and J. Speth, Phys. Rev. C {\bf  
    49}, 2671 (1994).

\bibitem{RJM84}R. J. Mcleod and D. J. Ernst, Phys. Rev. C {\bf 29}, 906, (1984).

\bibitem{Gieb}  D. R. Giebink. Phys. Rev. C {\bf 32}, 502 (1985);
    D. R. Giebink and D. J. Ernst, Comp. Phys. Comm. {\bf 48}, 407 (1988).

\bibitem{DVA92}D. V. Ahluwalia and D. J. Ernst, Phys. Rev. C {\bf 45}, 3010 (1992);
   D. J. Ernst, in {\it Lorentz Group, CPT, and Neutrinos}, ed.~A. E. Chubykalko, V. V.    
   Dvoeglazov, D. J. Ernst, V. G. Kadyshevsky, and Y. S. Kim (World Scientific, NY, 2000)  
   p.~47.

\bibitem{SBJ} S. B. Jacobson, Ph. D. thesis, Vanderbilt University, 2000            
   (unpublished).

\bibitem{Lee} T. D. Lee, Phys. Rev. {\bf 9}, 1329 (1954).

\bibitem{VGK81} V. G. Kadyshevski, Nucl. Phys. {\bf B6}, 125 (1981).
   
\bibitem{CL} G. Chew and F. Low, Phys. Rev. {\bf 101}, 1570 (1956).

\bibitem{low} F. Low, Phys. Rev. {\bf 97}, 1392 (1954);
    F. M. H. Villars, ``Collision Theory'' in 
    {\it Fundamentals Of Nuclear Theory}, (I. A. E. C, Vienna, 1967);
    E. F. Redish and F. M. H. Villars, Ann. Phys. (N.Y.) {\bf 56}, 225 (1970); 
    D. J. Ernst, Ph. D. thesis, M. I. T., 1970 (unpublished).    

\bibitem{RJM82} R. J. McLeod and D. J. Ernst, Phys. Lett. {\bf 119B}, 277 (1982);
    Nucl. Phys. {\bf A437}, 669 (1985).

\bibitem{chew}G. F. Chew, Phys Rev. {\bf 94}, 1748 (1954); {\bf 94}, 1755 (1954).

\bibitem{RA95} R. A. Arndt, I. I. Strakovsky, and R. L. Workman, Phys. Rev. C {\bf 52},
    2246 (1995).

\bibitem{DJE95} D.~J.~Ernst, in {\it $\Pi$N Newsletter, Proceedings of the Sixth           
   International Symposium on Meson-Nucleon Physics and the Structure of the Nucleon},
   ed. D.~Drechsel, G.~H\"oler, W. Kluge, and B.~M.~K.~Nefkins, {\bf 11},
   11 (1995).

\bibitem{MWRS} M. W. Rawool--Sullivan, C. L. Morris, J.M. O'Donnell, R. M. Whitton, B. K.  
   Park, G. R. Burleson, D. L. Watson, J. Johnson, A. L. Williams, D. A. Smith, D. J.      
   Ernst and C. M. Chen, Phys. Rev. C {\bf 49}, 627 (1994);
   G. Kahrimanis, G. Burleson, C. M. Chen, B. C. Clark, K. Dhuga, D. J. Ernst, J. A.       
   Faucett, H. T. Fortune, S. Hama, A. Hussein, M. F. Jiang, K. W. Johnson, L. Kurth, S.   
   Matthews, J. McGill, R. L. Mercer, C. F. Moore, S. Mordechai, C. L. Morris, J.          
   O'Donnell, M. Snell, M. Rawool-Sullivan, L. Ray, C. Whitley, and A. Williams, Phys.     
   Rev. C {\bf 55}, 2533 (1997).

   
\end{references}
\end{document}